# A Data-Based Approach to Social Influence Maximization

Amit Goyal
University of British Columbia
Vancouver, BC, Canada
goyal@cs.ubc.ca

Francesco Bonchi
Yahoo! Research
Barcelona, Spain
bonchi@yahoo-inc.com

Laks V. S. Lakshmanan
University of British Columbia
Vancouver, BC, Canada
laks@cs.ubc.ca

## ABSTRACT

Influence maximization is the problem of finding a set of users in a social network, such that by targeting this set, one maximizes the expected spread of influence in the network. Most of the literature on this topic has focused exclusively on the social graph, overlooking historical data, i.e., traces of past action propagations. In this paper, we study influence maximization from a novel data-based perspective. In particular, we introduce a new model, which we call *credit distribution*, that directly leverages available propagation traces to learn how influence flows in the network and uses this to estimate expected influence spread. Our approach also learns the different levels of influenceability of users, and it is time-aware in the sense that it takes the temporal nature of influence into account.

We show that influence maximization under the credit distribution model is **NP**-hard and that the function that defines expected spread under our model is submodular. Based on these, we develop an approximation algorithm for solving the influence maximization problem that at once enjoys high accuracy compared to the standard approach, while being several orders of magnitude faster and more scalable.

## 1. INTRODUCTION

Motivated by applications such as viral marketing [5], personalized recommendations [15], feed ranking [8], and the analysis of Twitter [16, 1], the study of the propagation of influence exerted by users of an online social network on other users has received tremendous attention in the last years. One of the key problems in this area is the identification of influential users, by targeting whom certain desirable outcomes can be achieved. Here, targeting could mean giving free (or price discounted) samples of a product and the desired outcome may be to get as many customers to buy the product as possible. Kempe *et al.* [10] formalized this as the *influence maximization* problem: find $k$ "seed" nodes in the network, for a given number $k$, such that by activating them we can maximize the *expected influence spread*, i.e., the expected number of nodes that eventually get activated, according to a chosen *propagation model*. The propagation model governs how influence diffuses or propagates through the network (see Section 2 for background on the most prominent propagation models adopted by [10]). Following this seminal paper, there has been substantial work in this area (see Section 2.1). In this paper we study influence maximization as defined by Kempe *et al.*, but from a novel, data-based perspective.

Influence maximization requires two kinds of data – a directed graph $G$ and an assignment of probabilities (or weights) to the edges of $G$, capturing degrees of influence. E.g., in Figure 1, the probability of the edge $(v,u)$ is 0.25 and it says there is a probability 0.25 with which user $v$ influences $u$ and thus $v$'s actions will propagate to $u$ with probability 0.25. In real life, while the digraph representing a social network is often explicitly available, edge probabilities are not. Facing difficulties in gathering real action propagation traces from which to "learn" edge probabilities, previous work has resorted to simply making assumptions about these probabilities. The methods adopted for assignment of probabilities to edges include the following: (i) treating them as constant (e.g., 0.01), (ii) drawing values uniformly at random from a small set of constants, e.g., $\{0.1, 0.01, 0.001\}$ in the so-called trivalency "model", or (iii) defining them to be the reciprocal of a node's in-degree, in the so-called weighted cascade "model" (see e.g., [10, 3, 2]). Only recently researchers have shown how to *learn* the edge probabilities from real data on past propagation traces of actions performed by users (nodes) [14, 7].

Given that there have been several ad hoc assumptions about probability assignment as well as recent techniques for learning edge probabilities from real data, some natural questions arise. What is the relative importance of the graph structure and the edge probabilities in the influence maximization problem? To what extent different methods of edge probability assignment accurately describe the influence propagation phenomenon? In particular, how do the various edge probability assignments considered in earlier literature compare with probabilities learned from real data when it comes to accurately predicting the expected influence spread? Learning edge probabilities from real data is prone to error either owing to noise in the data or to the inherent nature of mining these probabilities. How robust are solutions to influence maximization against such noise?

As we will discuss in the next section, the influence maximization process based on Monte Carlo (MC) simulation is computationally expensive, even when the edge probabilities are given as input. Having to learn these probabilities, from a large database of traces, only adds to the complexity. *Can we avoid the costly learning and simulation approach, and directly mine the available log of past action propagation traces to build a model of the spread of any given seed set?*

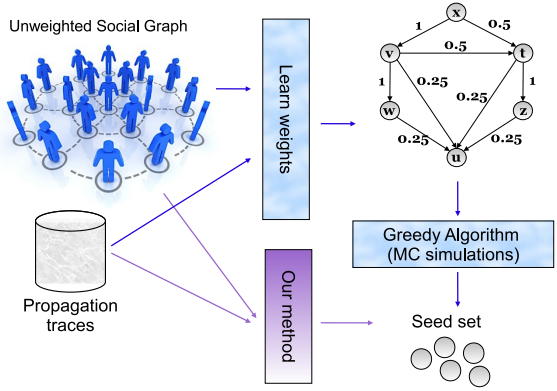

Figure 1: The standard influence maximization process (in light blue), and our approach (in magenta).

Our research is driven by the questions above, and it achieves the following contributions.

- We conduct a detailed empirical evaluation of different methods of edge probability assignment as well probabilities learned from real propagation traces and show that methods that don't learn probabilities from real data end up choosing very different seed sets than those that do. Secondly, we show the spread predicted by methods based on edge probability assignment suffers from large errors, compared to methods that learn edge probabilities from real data. This offers some evidence that the former class of methods risk choosing poor quality seeds (Section 3).

- We develop a new model called *credit distribution*, built on top of real propagation traces that allows us to directly predict the influence spread of node sets, without any need for learning edge probabilities or conducting MC simulations (Section 4).

- We show that influence maximization under credit distribution is **NP**-hard. However, we show the function defining influence spread under this model is monotone and submodular. Using this, we develop a greedy algorithm that guarantees a $(1-1/e)$-approximation to the optimal solution and is scalable (Section 5).

- We conduct a comprehensive set of experiments on large real-world datasets (Section 6). We compare our proposal against the standard approach of [10] with edge probabilities learned from real data, and show that the credit distribution model provides higher accuracy. We also demonstrate the scalability of our approach by showing our results on very large real world networks, on which the standard approach is not practical.

## 2. BACKGROUND

Given a directed graph $G = (V, E, p)$, where nodes are users and edges are labeled with influence probabilities among users, the influence maximization problem asks for a *seed set* of users, that maximizes the *expected spread* of influence in the social network, under a given propagation model. Kempe *et al.* [10] mainly focus on two propagation models – the *Independent Cascade* (IC) and the *Linear Threshold* (LT) models. In both, at a given time, each node

**Algorithm 1** Greedy

**Input:** $G, k, \sigma_m$
**Output:** seed set $S$
1: $S \leftarrow \emptyset$
2: **while** $|S| < k$ **do**
3: $\quad u \leftarrow \arg\max_{w \in V-S}(\sigma_m(S+w) - \sigma_m(S));$
4: $\quad S \leftarrow S + u$

can be either active or inactive. Each node's tendency to become active increases monotonically as more of its neighbors become active, and an active node never becomes inactive again. Time unfolds in discrete steps.

In the IC model, each active neighbor $v$ of a node $u$ has one shot at influencing $u$ and succeeds with probability $p_{v,u}$, the probability with which $v$ influences $u$. In the LT model, each node $u$ is influenced by each neighbor $v$ according to a weight $p_{v,u}$, such that the sum of incoming weights to $u$ is no more than 1. Each node $u$ chooses a threshold $\theta_u$ uniformly at random from $[0, 1]$. At any timestamp $t$, if the total weight from the active neighbors of an inactive node $u$ is at least $\theta_u$, then $u$ becomes active at timestamp $t + 1$.

In both the models, the process repeats until no new node becomes active. Given a propagation model $m$ (e.g., IC or LT) and an initial seed set $S \subseteq V$, the expected number of active nodes at the end of the process is the *expected (influence) spread*, denoted by $\sigma_m(S)$.

The *influence maximization problem* is defined as follows.

PROBLEM 1 (INFLUENCE MAXIMIZATION). *Given a directed and edge-weighted social graph $G = (V, E, p)$, a propagation model $m$, and a number $k \leq |V|$, find a set $S \subseteq V$, $|S| = k$, such that $\sigma_m(S)$ is maximum.*

Under both the IC and LT propagation models, this problem is shown to be **NP**-hard [10]. Kempe *et al.*, however, showed that the function $\sigma_m(S)$ is *monotone* and *submodular*. A function $f$ from sets to reals is monotone if $f(S) \leq f(T)$ whenever $S \subseteq T$. A function $f$ is submodular if $f(S + w) - f(S) \geq f(T + w) - f(T)$ whenever $S \subseteq T$.[1] Submodularity intuitively says an active node's probability of activating some inactive node $u$ does not increase if more nodes have already attempted to activate $u$.

For any monotone submodular function $f$ with $f(\emptyset) = 0$, the problem of finding a set $S$ of size $k$ such that $f(S)$ is maximum, can be approximated to within a factor of $(1-1/e)$ by a greedy algorithm [13], a result that directly carries over to the influence maximization problem [10] (see Algorithm 1).

The complex step of the greedy algorithm is in line 3, where we select the node that provides the largest marginal gain $\sigma_m(S+w) - \sigma_m(S)$ with respect to the expected spread of the current seed set $S$. Computing the expected spread given a seed set is #**P**-hard under both the IC model [2, 8] and the LT model [4]. In their paper, Kempe *et al.* run MC simulations of the propagation model for sufficiently many times (the authors report 10, 000 trials) to obtain an accurate estimate of the expected spread, resulting in a very long computation time.

In the majority of the literature on influence maximization following [10], the edge-weighted social graph is assumed as input to the problem, without addressing the question of how the probabilities are obtained. In Figure 1, we summarize the standard process followed in influence maximization and we make explicit the phase of learning the edge probabilities. The process starts with the (unweighted) social graph and a log of past action propagations that say when each user performed an action. The log is used to estimate influence probabilities among the nodes. This produces the directed edge-weighted graph which is then given as input to the greedy algorithm which produces the seed set using MC simulations.

---
[1]In the rest of the paper we write $S + w$ in place of $S \cup \{w\}$ and similarly $S - T$ in place of $S \setminus T$.

## 2.1 Other Related Work

Domingos and Richardson [5] first introduced the problem of identifying influential users for a marketing campaign as a learning problem, which Kempe *et al.* [10] subsequently formulated as an optimization problem. Exploiting submodularity, Leskovec *et al.* [12] develop an efficient algorithm called CELF, based on a "lazy-forward" optimization in selecting new seeds. CELF is up to 700 times faster than the simple greedy algorithm, while delivering the same approximation guarantee (more details in Section 5.3). In spite of this big improvement their method still faces serious scalability issues [3], which has motivated recent works on efficient heuristics for overcoming the efficiency and scalability limits of the greedy algorithm [11, 3, 2, 4].

Chen *et al.* [2] propose PMIA heuristic to estimate influence spread under the IC model. They consider the influence flow via Maximum Influence Paths (MIP) instead of shortest path [11]. An MIP between a pair of nodes $(v, u)$ is the path with the maximum propagation probability from $v$ to $u$. More recently, Chen *et al.* [4] propose a scalable heuristic called LDAG for the LT model. They construct local DAGs for each node and consider influence only within it. Computing expected spread over DAGs can be done in linear time while over general graphs it is #**P**-hard [4]. While the PMIA and LDAG heuristics don't offer theoretical guarantees, the authors show empirically that these solutions are quite close to those obtained using the corresponding greedy algorithm. A key distinction with our work is that our proposal offers a *scalable solution to influence maximization with an approximation guarantee.*

The above body of work assumes a weighted social graph as input and does not address *how* the edge probabilities may be obtained. Saito *et al.* [14] study how to learn the probabilities for the IC model from a set of past propagations. They formalize this as a likelihood maximization problem and then apply the expectation maximization (EM) algorithm to solve it. In our experiments, we use their method to learn the probabilities for the IC model.

Goyal *et al.* [7] also study the problem of learning influence probabilities. They focus on the time varying nature of influence, and on factors such as the influenceability of a specific user, and influence-proneness of a certain action. They also show that their methods can be used to predict whether a user will perform an action and at *what time*, with higher accuracy for users with higher influenceability scores.

Our work is different from all of the above in that we propose a method for learning a model for directly predicting the influence spread for a given node set, bypassing the need to learn edge probabilities and to run expensive MC simulations. We use this as a basis to develop a scalable approximation algorithm for influence maximization that does not make use of any explicit propagation model and is instead data-based. To the best of our knowledge, such a data-based approach to influence maximization is novel.

## 3. WHY DATA MATTERS

What is the relative importance of the network structure and the edge probabilities in determining influence propagation? How important is it to accurately learn probabilities from real propagation traces? We have seen that a large majority of the literature assumes edge probabilities to be randomly chosen from an arbitrary fixed set or to be determined by node degrees. How do these methods compare with that of learning edge probabilities from real data, in terms of the quality of seeds selected? To answer this, we compare the performance of Algorithm 1 under the IC model, with different methods of assigning edge probability. To this end, we present two kinds of experiments that, to the best of our knowledge, have never been reported before.

**Datasets.** We take two real world datasets: Flixster and Flickr, both consisting of an unweighted directed social graph, along with an associated *action log*. An action log is a set of triples $(u, a, t)$ which say user $u$ performed action $a$ at time $t$. We refer to the set of triples in the action log corresponding to a specific action $a$ as the *propagation trace* (propagation for short) associated with $a$. Flixster (www.flixster.com) is one of the main players in the mobile and social movie rating business [9]. Here, an action is a user rating a movie. In other words, if user $v$ rates "The King's Speech", and later on $v$'s friend $u$ does the same, we consider the action of rating "The King's Speech" as having propagated from $v$ to $u$. Flickr is a popular photo sharing platform. Here, an action is a user joining an interest group (e.g., *"Nikon Selfportrait", "HDR Panoramas"*). The raw versions of both datasets are very large and as a result, experiments that require repeated MC simulations cannot be run within any reasonable time on the full data set. While the large version of the datasets are useful for testing scalability of our proposal, for other experiments we have to sample smaller datasets. In what follows, we use samples that correspond to taking a unique "community", obtained by means of graph clustering performed using $Graclus^2$. The resulting datasets are named FLIXSTER_SMALL and FLICKR_SMALL (statistics in Table 1).

|  | FLIXSTER LARGE | FLICKR LARGE | FLIXSTER SMALL | FLICKR SMALL |
| --- | --- | --- | --- | --- |
| #Nodes | 1M | 1.32M | 13K | 14.8K |
| #Dir. Edges | 28M | 81M | 192.4K | 1.17M |
| Avg. degree | 28 | 61 | 14.8 | 79 |
| #propagations | 49K | 296K | 25K | 28.5K |
| #tuples | 8.2M | 36M | 1.84M | 478K |

Table 1: Statistics of datasets.

One of the goals of the experiments is to determine which method more accurately predicts the expected spread of node sets. So we split the action log into two sets of propagation traces – training and test sets. The edge probabilities are learnt from the training set and thus, it is crucial that the splitting is performed in such a way that a propagation trace in its entirety falls into training or test set. Taking care that similar distributions of propagation sizes are maintained in the two sets, we place 80% and 20% of the propagations in training and test set respectively. Precisely, we sorted the propagation traces based on their size and put every fifth propagation in this ranking in the test set. As a result, the number of propagations in the training set are 5.1K and 5.7K for FLIXSTER_SMALL and FLICKR_SMALL respectively. The number of tuples in the training set are 1.5M and 385.3K respectively. The training set is used to learn the edge probabilities according to the EM-based method of Saito et al. [14]. One issue in using their method is that in their work, Saito et al. assume that the input action log data is as though it was generated by an IC model: i.e., time is discrete, and if user $u$ activates at time $t$, then at least one of the neighbors of $u$ was activated at time $t-1$. In real-world propagations this is not the case. To close this gap between their model and the real data, we let all previously activated neighbors of a node be its possible influencers.

---

[2] http://www.cs.utexas.edu/users/dml/Software/graclus.html

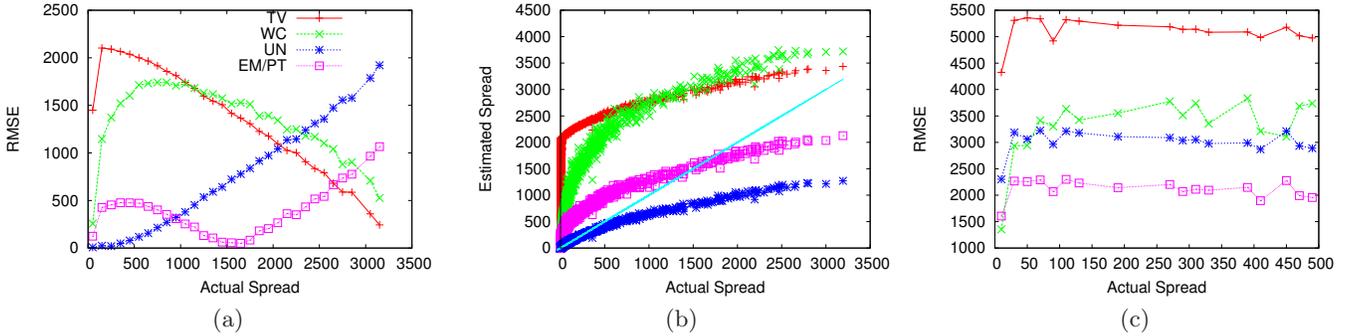

Figure 2: Error as a function of Actual Spread on (a) Flixster_Small, (c) Flickr_Small; (b) Scatter plot of predicted spread vs. actual spread on Flixster_Small. The legend in all the plots follows from (a).

**Methods experimented.** In both our experiments, we consider the IC model together with the following methods of edge probability assignment based on previous work [10, 14, 3, 2]:

**WC**: probability on an edge $(v, u)$ is 1/in-degree($u$) (known as weighted cascade);

**TV**: probabilities are selected uniformly at random from the set $\{0.1, 0.01, 0.001\}$ (trivalency).

**UN**: all edges are uniformly assigned probability $p = 0.01$.

**EM**: probabilities are learned from the training set using the EM-based method [14].

**PT**: Finally, in order to assess how robust the greedy method is to noise in the probability learning phase, we take EM-learnt probability and add noise. More precisely, for each edge $(v,u)$ we randomly pick a percentage from the interval $[-20\%, 20\%]$ to perturb $p_{v,u}$, rounding to 0 or 1 in cases that go below 0 or over 1 respectively. We call this method **PT** (EM perturbed).

**Experiment 1: Seed set intersection.** The goal of this experiment is to understand the extent to which choice of edge probabilities affects the decisions of different methods in seed set selection. We run Algorithm 1 under the IC model[3] with the various methods of probability assignment above, as well as with edge probabilities learned from the training data set using EM. In each case, we used the algorithm to produce a seed set of size $k = 50$. Table 2 reports the size of the intersection for each pair of seed sets. We can see that EM, the method using real data to learn the influence probabilities has a very small, almost empty, intersection with all other methods, with the exception of its own perturbed version PT. Thus, we conclude *all methods that use edge probabilities based on ad hoc assumptions select seed sets very different from the method that uses propagation trace data to learn edge probabilities. Secondly, noise in the learned edge probabilities does not affect the seed set selection too drastically, as shown by the intersection between EM and PT.* What can we say about the *quality of seed sets* chosen by the methods UN, TV, and WC, compared to EM? This is addressed next.

---

[3]We found the greedy algorithm on FLICKR_SMALL is too slow to complete in a reasonable time, even with CELF optimization. Hence we use the PMIA heuristic [2] (discussed in Section 2.1) in order to speed up IC computation. [2] empirically showed PMIA produces results very close to greedy.

| UN | WC | TV | EM | PT |    |    | PT | EM | TV | WC | UN |
|----|----|----|----|----|----|----|----|----|----|----|----|
| 50 | 25 | 5  | 6  | 6  | UN |    | 0  | 0  | 44 | 19 | 50 |
|    | 50 | 9  | 3  | 2  | WC |    | 0  | 0  | 17 | 50 |    |
|    |    | 50 | 3  | 2  | TV |    | 0  | 0  | 50 |    |    |
|    |    |    | 50 | 44 | EM |    | 44 | 50 |    |    |    |
| FLIXSTER_SMALL |    |    |    | 50 | PT |    | 50 |    |    | FLICKR_SMALL |  |

Table 2: Size of seed set intersection for $k = 50$ on Flixster_Small (left) and Flickr_Small (right).

**Experiment 2: Spread prediction.** In the second experiment, we address the question, how good each of the methods is at predicting the actual spread. For that end, for a given seed set $S$, we compute the expected spread $\sigma_{IC}(S)$ predicted by each of the methods and compare it with the actual spread of $S$ according to ground truth. For ground truth, for each propagation (movie in Flixster and group in Flickr) in the test set, we take the set of users that are the first to rate the movie (or join the group in case of Flickr) among their friends, i.e., the set of "initiators" of the action, to be the seed set. The actual spread is the number of users who performed that action, also called propagation size. This allows for a fair comparison of all methods from a neutral standpoint, which is a first in itself.

Figures 2(a) and (c) report the root mean squared error (RMSE) between predicted and actual spread on the two datasets: propagations in the test set are grouped in bins with respect to their size[4] and RMSE is computed inside each bin. On FLIXSTER_SMALL, uniform method (UN) works well but only for small propagations, and trivalency (TV) and weighted cascade (WC) work well but only for very large propagations (which are only few cases, i.e., outliers), and this is explainable with the fact that they *always* tend to predict the spread as very high. This is clearly shown in Figure 2(b), which shows a scatter plot between predicted and actual spread on FLIXSTER_SMALL.

---

[4]In FLIXSTER_SMALL bins are defined at multiples of 100, in FLICKR_SMALL at multiples of 20.

In the case of FLICKR_SMALL, EM clearly outperforms all the other methods for all sizes of actual spread (Figure 2(c)). In all cases, the performance of EM and PT are so close that they are almost indistinguishable.

Overall, even if EM tends to underestimate the spread when this gets larger, *it is by far the most accurate method with respect to the ground truth*. The other methods that do not use the real propagation traces to learn influence probabilities are found to be unreliable in predicting the true spread.

By putting the results of the two experiments together, we can draw the first conclusion of this paper: methods UN, TV, and WC select seed sets that are very different from EM and since they can be quite inaccurate in predicting the true spread, they can end up selecting seed sets of poor quality. It is thus extremely important to exploit available past propagation traces to learn the probabilities right. This finding strengthens the motivation for the rest of our work.

## 4. CREDIT DISTRIBUTION MODEL

The propagation models discussed in Section 2 are probabilistic in nature. In the IC model, coin flips decide whether an active node will succeed in activating its peers. In the LT model it is the node threshold chosen uniformly at random, together with the influence weights of active neighbors, that decides whether a node becomes active. Under both models, we can think of a propagation trace as a *possible world*, i.e., a possible outcome of a set of probabilistic choices.

Given a propagation model and a directed and edge-weighted social graph $G = (V, E, p)$, let $\mathbb{G}$ denote the set of all possible worlds. Independently of the model $m$ chosen, the expected spread $\sigma_m(S)$ can be written as:

$$\sigma_m(S) = \sum_{X \in \mathbb{G}} Pr[X] \cdot \sigma_m^X(S) \quad (1)$$

where $\sigma_m^X(S)$ is the number of nodes reachable from $S$ in the possible world $X$. The number of possible worlds is clearly exponential. Indeed, computing $\sigma_m(S)$ under the IC and LT models is #**P**-hard [2, 4], and the standard approach (see [10]) tackles influence spread computation from the perspective of Eq. (1): sample a possible world $X \in \mathbb{G}$, compute $\sigma_m^X(S)$, and repeat until the number of sampled worlds is large enough. We now develop an alternative approach for computing influence spread, by rewriting Eq. (1), giving a different perspective. Let $path(S, u)$ be an indicator random variable that is 1 if there exists a directed path from the set $S$ to $u$ and 0 otherwise. Moreover let $path_X(S, u)$ denote the outcome of the random variable in a possible world $X \in \mathbb{G}$. Then we have:

$$\sigma_m^X(S) = \sum_{u \in V} path_X(S, u) \quad (2)$$

Substituting in (1) and rearranging the terms we have:

$$\sigma_m(S) = \sum_{u \in V} \sum_{X \in \mathbb{G}} Pr[X] \, path_X(S, u) \quad (3)$$

From the definition of expectation, we can rewrite this to

$$\sigma_m(S) = \sum_{u \in V} E[path(S, u)] = \sum_{u \in V} Pr[path(S, u) = 1] \quad (4)$$

That is, the expected spread of a set $S$ is the sum over each node $u \in V$, of the probability of the node $u$ getting activated given that $S$ is the initial seed set.

The standard approach samples possible worlds from the perspective of Eq. (1). To leverage available data on real propagation traces, we observe that these traces are similar to possible worlds, except they are *"real available worlds"*. Thus, in this paper, we approach the computation of influence spread from the perspective of Eq. (4), i.e., we estimate directly $Pr[path(S, u) = 1]$ using the propagation traces that we have in the action log.

**Data Model.** We are given a social graph $G = (V, E)$, with nodes $V$ corresponding to users and directed (unweighted) edges $E$ corresponding to social ties between users, and an *action log*, i.e., a relation $\mathbb{L}(User, Action, Time)$ where a tuple $(u, a, t) \in \mathbb{L}$ indicates that user $u$ performed action $a$ at time $t$. It contains such a tuple for every action performed by every user of the system. We will assume that the projection of $\mathbb{L}$ on the first column is contained in the set of nodes $V$ of the social graph $G$. We let $\mathcal{A}$ denote the universe of actions, i.e., the projection of $\mathbb{L}$ on the second column. Moreover, we assume that a user performs an action at most once, and define the function $t(u, a)$ to return the time when user $u$ performed action $a$ (the value of $t(u, a)$ is undefined if $u$ never performed $a$, and $t(u, a) < t(v, a)$ is false whenever either of $t(u, a), t(v, a)$ is undefined).

We say that $a$ propagates from node $u$ to $v$ iff $u$ and $v$ are socially linked, and $u$ performs $a$ before $v$ (we also say that $u$ influences $v$ on $a$). This defines a *propagation graph* of $a$ as a directed graph $G(a) = (V(a), E(a))$, with $V(a) = \{v \in V \mid \exists t : (v, a, t) \in \mathbb{L}\}$ and $E(a) = \{(u, v) \in E \mid t(u, a) < t(v, a)\}$. Note that the propagation graph of an action $a$ is the graph-representation of the propagation trace of $a$, and it is always a DAG: it is directed, each node can have zero or more parents, and cycles are impossible due to the time constraint. The action log $\mathbb{L}$ is thus a set of these DAGs representing propagation traces through the social graph. We denote by $N_{in}(u, a) = \{v \mid (v, u) \in E(a)\}$ the set of *potential influencers* of $u$ for action $a$ and $d_{in}(u, a) = |N_{in}(u, a)|$ to be the *in-degree* of $u$ for action $a$. Finally, we call a user $u$ an *initiator* of action $a$ if $u \in V(a)$ and $d_{in}(u, a) = 0$, i.e., $u$ performed action $a$ but none of its neighbors performed it before $u$ did. Table 3 summarizes the notation used.

**The Sparsity Issue.** In order to estimate $Pr[path(S, u) = 1]$ using available propagation traces, it is natural to interpret such quantity as the fraction of the actions initiated by $S$ that propagated to $u$, given that $S$ is the seed set. More precisely, we could estimate this probability as

$$\frac{|\{a \in \mathcal{A} | initiate(a, S) \,\&\, \exists t : (u, a, t) \in \mathbb{L}\}|}{|\{a \in \mathcal{A} | initiate(a, S)\}|}$$

where $initiate(a, S)$ is true iff $S$ is precisely the set of initiators of action $a$. Unfortunately, this approach suffers from a *sparsity issue* which is intrinsic to the influence maximization problem [10]. If we need to be able to estimate $Pr[path(S, u) = 1]$ for any set $S$ and node $u$, we will need an enormous number of propagation traces corresponding to various combinations, where each trace has as its initiator set precisely the required node set $S$. It is clearly impractical to find a real action log where this can be realized. To overcome this obstacle, we propose a different approach to estimating $Pr[path(S, u) = 1]$ by taking a *"u-centric"* perspective: we assign "credits" to the possible influencers of a node $u$ whenever $u$ performs an action. The model is formally described next.

| | |
|---|---|
| $\mathcal{A}_u$ | Number of actions performed by $u$. |
| $N_{\text{in}}(u, a)$ | Neighbors of $u$ which activated on action $a$ before i.e., $u$'s potential influencers on action $a$. |
| $\gamma_{v,u}(a)$ | Direct influence credit given to $v$ for influencing $u$ for action $a$. |
| $\Gamma_{v,u}(a)$ | Total credit given to $v$ for influencing $u$ for action $a$. |
| $\kappa_{v,u}$ | Total credit given to $v$ for influencing $u$ for all actions. |
| $\Gamma^W_{x,u}(a)$ | Total credit given to $x$ for influencing $u$ for action $a$ considering the paths that are completely contained in $W \subseteq V$. |
| $\tau_{v,u}$ | Average time taken by actions to propagate from user $u$ to user $v$. |

**Table 3: Notation adopted in the next sections.**

**Credit Distribution.** When a user $u$ performs an action $a$, we want to give *direct influence credit*, denoted by $\gamma_{v,u}(a)$, to all $v \in N_{\text{in}}(u, a)$, i.e., all neighbors of $u$ that have performed the same action $a$ before $u$. We constrain the sum of the direct credits given by a user to its neighbors to be no more than 1. We can have various ways of assigning direct credit: for ease of exposition, we assume for the moment to give equal credits to each neighbor $v$ of $u$, i.e., $\gamma_{v,u}(a) = 1/d_{\text{in}}(u, a)$ for all $v \in N_{\text{in}}(u, a)$. Later we will see a more sophisticated method of assigning direct credit.

Intuitively, we also want to distribute influence credit transitively backwards in the propagation graph $G(a)$, such that not only $u$ gives credit to the users $v \in N_{\text{in}}(u, a)$, but they in turn pass on the credit to their predecessors in $G(a)$ and so on. This suggests the following definition of *total credit* given to a user $v$ for influencing $u$ on action $a$, corresponding to multiple propagation paths:

$$\Gamma_{v,u}(a) = \sum_{w \in N_{\text{in}}(u,a)} \Gamma_{v,w}(a) \cdot \gamma_{w,u}(a) \qquad (5)$$

where the base of the recursion is $\Gamma_{v,v}(a) = 1$. Sometimes, when the action is clear from the context, we can omit it and simply write $\gamma_{v,u}$ and $\Gamma_{v,u}$. From here on, as a running example, we consider the influence graph in Figure 1 as the propagation graph $G(a)$ with edges labeled with direct credits $\gamma_{v,u}(a) = 1/d_{in}(u,a)$. For instance,

$$\Gamma_{v,u} = \Gamma_{v,v} \cdot \gamma_{v,u} + \Gamma_{v,t} \cdot \gamma_{t,u} + \Gamma_{v,w} \cdot \gamma_{w,u} + \Gamma_{v,z} \cdot \gamma_{z,u}$$
$$= 1 \cdot 0.25 + 0.5 \cdot 0.25 + 1 \cdot 0.25 + 0.5 \cdot 0.25 = 0.75.$$

We next define the total credit given to a set of nodes $S \subseteq V(a)$ for influencing user $u$ on action $a$ as follows:

$$\Gamma_{S,u}(a) = \begin{cases} 1 & \text{if } v \in S; \\ \sum_{w \in N_{\text{in}}(u,a)} \Gamma_{S,w}(a) \cdot \gamma_{w,u}(a) & \text{otherwise} \end{cases}$$

Consider again the propagation graph $G(a)$ in Figure 1. Let $S = \{v, z\}$. Then, $\Gamma_{S,u}$ is the fraction of flow reaching $u$ that flows from either $v$ or $z$:

$$\Gamma_{S,u} = \Gamma_{S,w} \cdot \gamma_{w,u} + \Gamma_{S,v} \cdot \gamma_{v,u} + \Gamma_{S,t} \cdot \gamma_{t,u} + \Gamma_{S,z} \cdot \gamma_{z,u}$$
$$= 1 \cdot 0.25 + 1 \cdot 0.25 + 0.5 \cdot 0.25 + 1 \cdot 0.25 = 0.875.$$

**Aggregating Over All Actions and All Nodes.** The next question is how to aggregate the influence credit over the whole action log $\mathbb{L}$. Consider two nodes $v$ and $u$: the total influence credit given to $v$ by $u$ for all actions in $\mathcal{A}$, is simply obtained by taking the total credit over all actions and normalizing it by the number of actions performed by $u$ (denoted $\mathcal{A}_u$). This is justified by the fact that credits are assigned by $u$ backward to its potential influencers. We define:

$$\kappa_{v,u} = \frac{1}{\mathcal{A}_u} \sum_{a \in \mathcal{A}} \Gamma_{v,u}(a) \qquad (6)$$

Intuitively, it denotes the average credit given to $v$ for influencing $u$, over all actions that $u$ performs. Similarly, for the case of a set of nodes $S \subseteq V$, we can define the total influence credit for all the actions in $\mathcal{A}$ as:

$$\kappa_{S,u} = \frac{1}{\mathcal{A}_u} \sum_{a \in \mathcal{A}} \Gamma_{S,u}(a) \qquad (7)$$

Note that $\kappa_{S,u}$ corresponds, in our approach, to $Pr[path(S, u) = 1]$ in Eq. 4. Finally, inspired by Eq. 4, we define the influence spread $\sigma_{cd}(S)$ as the total influence credit given to $S$ from the whole social network:

$$\sigma_{cd}(S) = \sum_{u \in V} \kappa_{S,u} \qquad (8)$$

In the spirit of influence maximization (Problem 1), this is the objective function that we want to maximize. In the next section we formally state the problem of maximizing influence under the CD model. We prove that the problem is **NP**-hard and that the function $\sigma_{cd}(.)$ is submodular, paving the way for an approximation algorithm.

**Assigning Direct Credit.** We now revisit the problem of defining the direct credit $\gamma_{v,u}(a)$ given by a node $u$ to a neighbor $v$ for action $a$. In our previous work [7], we observed that influence decays over time in an exponential fashion and that some users are more influenceable than others. Motivated by these ideas, we propose to assign direct credit as:

$$\gamma_{v,u}(a) = \frac{infl(u)}{N_{\text{in}}(u,a)} \cdot \exp\left(-\frac{t(u,a) - t(v,a)}{\tau_{v,u}}\right) \qquad (9)$$

Here, $\tau_{v,u}$ is the average time taken for actions to propagate from user $v$ to user $u$. The exponential term in the equation achieves the desired effect that influence decays over time. Moreover, $infl(u)$ denotes the user influenceability, that is, how prone the user $u$ is to influence by the social context [7]. Precisely, $infl(u)$ is defined as the fraction of actions that $u$ performs under the influence of at least one of its neighbors, say $v$, i.e., $u$ performs the action, say $a$, such that $t(u, a) - t(v, a) \leq \tau_{v,u}$; this is normalized by $N_{\text{in}}(u, a)$ to ensure that the sum of direct credits assigned to neighbors of $u$ for action $a$ is at most 1. Note that both $infl(u)$ and $\tau_{\cdot,u}$ are learnt from (the training subset of) $\mathbb{L}$.

**Discussion.** It should be pointed out that unlike classical models such as IC and LT, the credit distribution model is *not* a propagation model. Instead, it is a model that, based on available propagation data, learns the total influence credit accorded to a given set $S$ by any node $u$ and uses this to predict the influence spread of $S$. It is not susceptible to the sparsity issue discussed above, and it obviates the need to perform expensive MC simulations for the purpose of estimating influence spread.

## 5. INFLUENCE MAXIMIZATION

We next formally define the problem studied in this paper.

PROBLEM 2 (INFLUENCE MAXIMIZATION - CD MODEL). *Given a directed social graph $G = (V, E)$, an action log $\mathbb{L}$, and an integer $k \leq |V|$, find a set $S \subseteq V$, $|S| = k$, that maximizes $\sigma_{cd}(S)$.*

THEOREM 1. *Influence maximization under the credit distribution model is* **NP**-*hard.*

PROOF. We prove the hardness by reducing the well-known **NP**-complete problem Vertex Cover [6] to our problem. Given an instance $\mathcal{I}$ of Vertex Cover, consisting of an undirected graph $G = (V, E)$ and a number $k$, create an instance $\mathcal{J}$ of the influence maximization problem under CD as follows. The directed social graph $G' = (V, E')$ associated with $\mathcal{J}$ has the same node set as $G$. $E'$ two directed edges $(u, v)$ and $(v, u)$ in place of every undirected edge $(u, v) \in E$. We express the action log $\mathbb{L}$ associated with $\mathcal{J}$ in terms of propagation graphs, for convenience. For each edge $(v, u) \in E$, create two propagation graphs, corresponding to two actions $a_1$ and $a_2$ in $\mathbb{L}$, consisting of only two nodes $v$ and $u$. In the propagation graph $G(a_1)$, create an edge from $v$ to $u$ indicating that the corresponding action is being propagated from $v$ to $u$. In the propagation graph $G(a_2)$, create an edge from $u$ to $v$. Assign direct credits $\gamma_{v,u}(a_1) = \gamma_{u,v}(a_2) = \alpha$ where $\alpha \in (0, 1]$. For instance, if we assign direct credits simply as $\gamma_{v,u}(a) = 1/d_{\text{in}}(u, a)$, then $\alpha = 1$. Similarly, if we assign direct credits as in equation 9, then $\alpha = 1/e$. The reduction clearly takes polynomial time. We next prove that a set $S \subseteq V$, with $|S| \leq k$, is a vertex cover of $G$ if and only if its influence spread $\sigma_{cd}(S)$ in the instance $\mathcal{J}$ is at least $k + \alpha \cdot (|V| - k)/2$.

**Only if:** Suppose $S$ is a vertex cover of $G$ in the instance $\mathcal{I}$. Consider any arbitrary node $u$. If $u \in S$, then $\kappa_{S,u} = 1$ by definition. On the other hand, if $u \notin S$, then $\kappa_{S,u} = \sum_a \Gamma_{S,u}(a)/(2 \cdot deg(u))$. Since $u$ is not in the vertex cover, all its neighbors must be in $S$ and thus, for exactly half of the actions $a$ that $u$ performs, $\Gamma_{S,u}(a) = \alpha$. These are the actions that $u$ performs after its neighbor. Hence, $\sum_a \Gamma_{S,u}(a) = \alpha \cdot deg(u)$ and $\kappa_{S,u} = \alpha/2$, where $deg(u)$ is $u$'s degree in $G$. This implies $\sigma_{cd}(S) = \sum_{u \in V} \kappa_{S,u} = k + \alpha \cdot (|V| - k)/2$.

**If:** Let $S$ be any seed set whose spread is at least $k + \alpha \cdot (|V| - k)/2$ in instance $\mathcal{J}$. Let $N(u)$ be the set of neighbors of $u$ in $G$, that is, $N(u) = \{v \in V | (v, u) \in E\}$. Consider an arbitrary node $u \notin S$. For each node $v$ in $N(u) \cap S$, $v$ has a credit of $\alpha$ over $u$, for the unique action whose propagation graph is the edge $(v, u)$, and a null credit for all the other actions. Therefore $\kappa_{v,u} = \alpha/(2 \cdot deg(u))$. Aggregating over the whole seed set $S$ we have that $\kappa_{S,u} = \alpha \cdot |N(u) \cap S|/(2 \cdot deg(u))$. Hence, $\sigma_{cd}(S) = \sum_{u \in V} \kappa_{S,u} = k + \sum_{u \in V \setminus S} \alpha \cdot |N(u) \cap S|/(2 \cdot (deg(u)))$.

From our assumption, $\sigma_{cd}(S) \geq k + \alpha \cdot (|V| - k)/2$, it follows that $\sum_{u \in V \setminus S} |N(u) \cap S|/deg(u) \geq |V| - k$. As $|N(u) \cap S| \leq deg(u)$, this is possible only when $\forall u \in V \setminus S : |N(u) \cap S| = deg(u)$, implying that all neighbors of $u$ must be in $S$ and therefore, $S$ is a vertex cover in instance $\mathcal{I}$. □

Since the problem is **NP**-hard, we are interested in developing an approximation algorithm. We prove that the influence spread function is submodular under the CD model, paving the way for efficient approximation.

THEOREM 2. $\sigma_{cd}(S)$ *is monotone and submodular.*

PROOF. It suffices to show that $\Gamma_{S,u}(a)$ is monotone and submodular as a positive linear combination of monotone, submodular functions is also monotone and submodular [13]. Clearly it is monotone. We prove submodularity by induction on path lengths. Note that the propagation graph for a given action is acyclic and hence the maximum path length is $|V| - 1$. Let $\Gamma_{S,u}(a, \ell)$ denote the total credit obtained by the set $S$ for influencing $u$, restricting attention to paths of length $\leq \ell$. Thus, $\Gamma_{S,u}(a) = \Gamma_{S,u}(a, |V| - 1)$.

Let $S$ and $T$ be two node sets such that $S \subseteq T$ and let $x \notin T$. Recall that the function $\Gamma$ is submodular iff $\Gamma_{S+x,u}(a) - \Gamma_{S,u}(a) \geq \Gamma_{T+x,u}(a) - \Gamma_{T,u}(a)$. We call the left hand side of the inequality the marginal gain of $x$ with respect to $S$ (implicitly understood to be on $u$) and similarly for the right hand side.

**Base Case:** In the base case, $\ell = 0$. Depending on $u$, the base case can be split into various sub-cases: (a) If $u \in S$, then the marginal gain of $x$ with respect to both $S$ and $T$ is 0; (b) If $u \in T, u \notin S$, then while $x$'s marginal gain with respect to $T$ is 0, its marginal gain with respect to $S$ is no less than 0 as the function $\Gamma(\cdot)$ is monotone; (c) If $u = x$, then the marginal gain of $x$ with respect to both $S$ and $T$ is exactly 1; (d) If $u \neq x$ and $u \notin T$, then the total credits on $u$ from $S$, $S + x$, $T$ and $T + x$ are 0. This proves the base case.

**Induction Step:** Assume that the function $\Gamma$ is submodular when restricted to path lengths $\leq \ell$, that is, $\forall w \in V$:

$$\Gamma_{S+x,w}(a, \ell) - \Gamma_{S,w}(a, \ell) \geq \Gamma_{T+x,w}(a, \ell) - \Gamma_{T,w}(a, \ell) \quad (10)$$

We will prove that the function remains submodular for paths of length $\ell + 1$ for any node $u \in V$. Consider the marginal gain of $x$ with respect to $S$ on $u$ when restricted to paths of length $\leq \ell + 1$, that is, consider $\Gamma_{S+x,u}(a, \ell+1) - \Gamma_{S,u}(a, \ell+1)$. By definition, this is equal to $\sum_{w \in N_{\text{in}}(u,a)} \Gamma_{S+x,w}(a, \ell) \cdot \gamma_{w,u}(a) - \sum_{w \in N_{\text{in}}(u,a)} \Gamma_{S,w}(a, \ell) \cdot \gamma_{w,u}(a)$. Taking the common factor $\gamma_{w,u}(a)$ out and applying induction hypothesis (Eq. 10), this is

$$\geq \sum_{w \in N_{\text{in}}(u,a)} (\Gamma_{T+x,w}(a, \ell) - \Gamma_{T,w}(a, \ell)) \cdot \gamma_{w,u}(a)$$
$$= \Gamma_{T+x,u}(a, \ell+1) - \Gamma_{T,u}(a, \ell+1)$$

This was to be shown. □

In the remaining sub-sections, we develop an efficient approximation algorithm to solve the influence maximization problem under the CD model.

## 5.1 Overview of our method

In the previous section, we show that while influence maximization under the CD model is **NP**-hard, the influence spread function is monotone and submodular. Consequently, the greedy algorithm (Algorithm 1) provides a $(1 - 1/e)$-approximation to the optimum solution [13]. However, the greedy algorithm by itself does not guarantee efficiency as it requires to compute the marginal gain of a candidate seed node with respect to the current seed set, i.e., $\sigma_{cd}(S + w) - \sigma_{cd}(S)$ (line 3 of Algorithm 1). For the IC and LT models, this is done by expensive MC simulations. For the CD model, the marginal gain can be directly computed from the action log $\mathbb{L}$. A naive way to do this would be to scan $\mathbb{L}$ in each iteration. But this approach would be very inefficient. Hence, we focus our attention on computing the marginal gain efficiently by carefully exploiting some properties of our model.

From here on, with a superscript $W \subseteq V$ on the function $\Gamma(\cdot)$, we denote the function to be evaluated on the sub-graph induced by nodes in $W$. For example, $\Gamma^W_{x,u}(a)$ is the total credit given to node $x$ for influencing node $u$ to perform action $a$ considering the paths that are contained completely in the sub-graph induced by $V(a) \cap W$. That is, the sub-graph of the propagation graph for action $a$, induced by the nodes $W \cap V(a)$. When the superscript is not present the graph considered is the whole propagation graph for action $a$, i.e., $\Gamma_{x,u}(a) = \Gamma_{x,u}^{V(a)}(a)$. It should be noted that the direct credit $\gamma_{x,u}$ is always assigned considering the whole propagation graph. The following result is key to the efficiency of our algorithm.

THEOREM 3.
$$\sigma_{cd}(S+x) - \sigma_{cd}(S) = \sum_{a \in \mathcal{A}} \left( (1 - \Gamma_{S,x}(a)) \cdot \sum_{u \in V} \frac{1}{\mathcal{A}_u} \cdot \Gamma_{x,u}^{V-S}(a) \right)$$

Intuitively, the theorem says that the marginal gain of a node $x$ equals the sum of *normalized* marginal gain of $x$ on all actions. We give more insights into this equation in the proof. The theorem provides us an efficient method to compute the marginal gain, given values of $\Gamma_{S,x}(a)$ and $\Gamma_{x,u}^{V-S}(a)$: this is the key idea behind the efficiency of our algorithm, which can be abstractly summarized as follows:

1. Initially, scan the action log $\mathbb{L}$ and compute $\Gamma_{v,u}(a)$ for all combinations of $v$, $u$ and $a$ (Algorithm 2). Note that at the beginning, $S = \emptyset$ and hence $\Gamma_{S,x}(a) = 0$ for all combinations of $x$ and $a$.

2. In each iteration of the greedy method, a node that provides the maximum marginal gain is added to the seed set. For this step we adopt the CELF [12] optimization idea (Algorithm 3).

3. To compute the marginal gain of a node $x$ efficiently, we use Theorem 3. It requires values of $\Gamma_{v,u}^{V-S}(a)$ and $\Gamma_{S,x}(a)$ (Algorithm 4).

4. Once a node is added to the seed set, $\Gamma_{v,u}^{V-S}(a)$ and $\Gamma_{S,x}(a)$ are updated using Lemmas 2 and 3 (Alg. 5).

### 5.2 Proof of Theorem 3

The proof of Theorem 3 is non-trivial and we need to prove a few auxiliary claims first. In the process, we also derive equations to update the total credit (step 4 in the outline of our algorithm above).

LEMMA 1. $\Gamma_{S,u}(a) = \sum_{v \in S} \Gamma_{v,u}^{V-S+v}(a)$

We first explain the claim by means of an example by taking the influence graph in Figure 1 as a propagation graph $G(a)$ with direct credit $\gamma_{v,u}(a) = 1/d_{in}(u,a)$. Let $S = \{v, z\}$, according to Lemma 1 (dropping the argument $a$), $\Gamma_{S,u} = \Gamma_{v,u}^{V-z} + \Gamma_{z,u}^{V-v} = (0.25 + 0.25 + 0.5 \cdot 0.25) + 0.25 = 0.875$. Note that the credit given to $v$ via the path $v \to t \to z \to u$ is ignored. Next, we formally prove the claim.

*Proof of Lemma 1:* By induction on path lengths. Recall that $\Gamma_{S,u}(a,l)$ denotes the total credit given to set $S$ for influencing node $u$ over paths of length no more than $l$.

**Base Case:** $l = 0$ implies only $u$ can get credit for influencing itself. Therefore if $u \notin S$, both sides of the equality become 0. When $u \in S$, the total credit given to $S$ for influencing $u$ (left hand side) is 1 by definition, while in the right hand side all terms in the summation are 0 except the case $v = u$, that is $\Gamma_{u,u}^{V-S+u}(a,0) = 1$.

**Induction Step:** Assume that the lemma is true for path lengths no more than $l$. We prove it for path length up to $l + 1$. We start with definition of $\Gamma_{S,u}(a, l+1)$ (Eq. 5). $\Gamma_{S,u}(a, l+1) = \sum_{w \in N_{in}(u,a)} \Gamma_{S,w}(a,l) \cdot \gamma_{w,u}(a)$. Applying induction hypothesis, the right hand side becomes:

$$= \sum_{w \in N_{in}(u,a)} \left( \sum_{v \in S} \Gamma_{v,w}^{V-S+v}(a,l) \right) \cdot \gamma_{w,u}(a) =$$

$$\sum_{v \in S} \left( \sum_{w \in N_{in}(u,a)} \Gamma_{v,w}^{V-S+v}(a,l) \cdot \gamma_{w,u}(a) \right) = \sum_{v \in S} \Gamma_{v,u}^{V-S+v}(a, l+1)$$

This concludes the proof. $\square$

Next, we show how the total credit can be updated incrementally, when the induced sub-graph under consideration changes. Consider the sub-graph induced by the nodes $W = V - S$ where $S$ is the current seed set. Let $\Gamma_{v,u}^{W}(a)$ be the total credit given to node $v$ for influencing $u$ in this sub-graph. Suppose node $x$ is added to the seed set, then we are interested in computing $\Gamma_{v,u}^{W-x}(a)$. This is clearly the total credit given to $v$ minus the total credit given to $v$ via paths that go through $x$. More precisely, we have:

LEMMA 2. $\Gamma_{v,u}^{W-x}(a) = \Gamma_{v,u}^{W}(a) - \Gamma_{v,x}^{W}(a) \cdot \Gamma_{x,u}^{W}(a)$.

As an example, consider again the propagation graph in Figure 1 with $S = \{t, z\}$. The total credit given to $v$ for influencing $u$ on the subgraph induced by nodes in $V - \{t, z\}$ is $1 \cdot 0.25 + 0.25 = 0.5$. Suppose $w$ is added to the seed set $S$, then $\Gamma_{v,u}^{V-S-w} = 0.5 - 1 \cdot 0.25 = 0.25$.

The next lemma shows how to update incrementally the total credit of influence given to a set $S$ by a node $u$, after $x$ is added to the set. This is needed in step 4 of our method as sketched in previous section.

LEMMA 3. $\Gamma_{S+x,u}(a) = \Gamma_{S,u}(a) + \Gamma_{x,u}^{V-S} \cdot (1 - \Gamma_{S,x}(a))$

*Proof:* Since we refer to a single action, we drop the argument $a$ and assume it implicitly. We use Lemma 1 to expand $\Gamma_{S+x,u}$ and $\Gamma_{S,u}$:

$$\Gamma_{S+x,u} - \Gamma_{S,u} = \sum_{v \in S+x} \Gamma_{v,u}^{V-S-x+v} - \sum_{v \in S} \Gamma_{v,u}^{V-S+v}$$

$$= \Gamma_{x,u}^{V-S} - \sum_{v \in S} (\Gamma_{v,u}^{V-S+v} - \Gamma_{v,u}^{V-S-x+v})$$

Applying Lemma 2 to the terms inside the summation (with $W = V - S + v$), the right hand side becomes

$$= \Gamma_{x,u}^{V-S} - \sum_{v \in S} (\Gamma_{v,x}^{V-S+v} \cdot \Gamma_{x,u}^{V-S+v})$$

The terms inside the summation denote the total credit given to $v$ for influencing $u$ considering the paths that go through $x$ in the sub-graph induced by the nodes $V - S + v$. Since the graph is acyclic, if $\Gamma_{v,x}^{V-S+v}$ is non-zero, then any path from $x$ to $u$ cannot pass through $v$. Hence, $\Gamma_{v,x}^{V-S+v} \cdot \Gamma_{x,u}^{V-S+v} = \Gamma_{v,x}^{V-S+v} \cdot \Gamma_{x,u}^{V-S}$. Note that this equality holds even when $\Gamma_{v,x}^{V-S+v} = 0$ as both sides would be 0 in that case. Thus, $\Gamma_{S+x,u} - \Gamma_{S,u} = \Gamma_{x,u}^{V-S} - \sum_{v \in S} (\Gamma_{v,x}^{V-S+v} \cdot \Gamma_{x,u}^{V-S})$.

The term $\Gamma_{x,u}^{V-S}$ can be taken out of the summation and applying Lemma 1 gives $\sum_{v \in S} \Gamma_{v,x}^{V-S+v} = \Gamma_{S,x}$, from which the lemma follows. $\square$

Finally, we are ready to prove Theorem 3.

*Proof of Theorem 3:* By definition,

$$\sigma_{cd}(S+x) - \sigma_{cd}(S) = \sum_{u \in V} \frac{1}{\mathcal{A}_u} \sum_{a \in \mathcal{A}} (\Gamma_{S+x,u}(a) - \Gamma_{S,u}(a))$$

Applying lemma 3, the right hand side becomes

$$= \sum_{u \in V} \frac{1}{\mathcal{A}_u} \sum_{a \in \mathcal{A}} (\Gamma_{x,u}^{V-S}(a) \cdot (1 - \Gamma_{S,x}(a)))$$

Rearranging the terms, we get $\sigma_{cd}(S+x) - \sigma_{cd}(S) = \sum_{a \in \mathcal{A}} \left( (1 - \Gamma_{S,x}(a)) \cdot \sum_{u \in V} \frac{1}{\mathcal{A}_u} \cdot \Gamma_{x,u}^{V-S}(a) \right)$, which was to be shown. □

## 5.3 Algorithms

In this section, we present our algorithm which builds on the properties developed in previous sections and whose outline was given in Section 5.1. Initially, we scan the action log $\mathbb{L}$ and then, we use the greedy algorithm with CELF optimization to select the seed set. While scanning the action log, we maintain all the information needed to select $k$ seeds later. In particular, we compute total credit given to each node $v$ for influencing any other node $u$ for all actions $a$ and record it into the data structure $UC$ (User Credits). Each entry $UC[v][u][a]$ corresponds to $\Gamma_{v,u}^{V-S}(a)$, that is, total credit given to $v$ for activating $u$ on the graph induced by $V - S$ where $S$ is the current seed set. We also maintain another data structure $SC$ (Set Credits) where each entry $SC[x][a]$ refers to the total credit given to the current seed set $S$ by a node $x$ for an action $a$, that is, $\Gamma_{S,x}(a)$. Since $S$ is empty in the beginning, $SC$ is not used in the first iteration.

---
**Algorithm 2** *Scan*

**Input:** $G, \mathbb{L}, \lambda$
**Output:** $UC$
1: $UC \leftarrow \emptyset$
2: **for** each action $a$ in $\mathbb{L}$ **do**
3:    $current\_table \leftarrow \emptyset$
4:    **for** each tuple $\langle u, a, t_u \rangle$ in chronological order **do**
5:      $Parents(u) \leftarrow \emptyset$; $\mathcal{A}_u \leftarrow \mathcal{A}_u + 1$; $UC[*][u][a] \leftarrow 0$
6:      **while** $\exists v : (v, u) \in G, v \in current\_table$ **do**
7:        $Parents(u) \leftarrow Parents(u) \cup \{v\}$
8:      **for** each $v \in Parents(u)$ **do**
9:        compute $\gamma_{v,u}$
10:       **if** $\gamma_{v,u} \geq \lambda$ **then**
11:         $UC[v][u][a] \leftarrow UC[v][u][a] + \gamma_{v,u}$
12:         **for** each $w$ such that $UC[w][v][a] \cdot \gamma_{v,u} \geq \lambda$ **do**
13:           $UC[w][u][a] \leftarrow UC[w][u][a] + \gamma_{v,u} UC[w][v][a]$
14:    $current\_table \leftarrow current\_table \cup \{u\}$

---

Algorithm 2 describes the first step of our method that scans $\mathbb{L}$. $\mathbb{L}$ is maintained sorted, first by action and then by time. It processes one action at a time and in chronological order. We use *current_table* to maintain the list of users who have performed the current action and have been seen so far, and $\mathcal{A}_u$ to denote the number of actions performed by user $u$ in $\mathbb{L}$, and $Parents(u)$ for the list of parents of each user $u$ with respect to the current action $a$, that is, $N_{\text{in}}(u, a)$. For each action $a$ and for each user $u$ that performs it, we scan *current_table* to find its neighbors that already performed $a$ and add them to the list of parents of $u$. Then for each parent $v$ of $u$, we compute the direct credit $\gamma_{v,u}(a)$ appropriately (line 9). For the ease of exposition, here we assume the simple definition $\gamma_{v,u}(a) = 1/N_{\text{in}}(u, a)$, that can be implemented as $\gamma = 1/|Parents(u)|$. If we want to use the more complex definition of direct credit given in Eq. (9), we need to learn the parameters $\tau_{v,u}$ for all edges and $\textit{infl}(u)$ for all nodes in advance and pass them on to Algorithm 2 as input, similarly to what the standard method does for influence probabilities. Although it is straightforward to learn these parameters by means of a preliminary scan of $\mathbb{L}$, we refer the reader to [7] for an efficient way to learn them. The total credit given to various nodes for influencing $u$ is then computed using equation 5 (lines 10-13). For the sake of reducing memory requirements, we use a truncation threshold $\lambda$ and discard credits that are below the threshold. In the experiments we will assess the effect of this truncation.

---
**Algorithm 3** *Greedy* with CELF

**Input:** $UC, k$
**Output:** seed set $S$
1: $SC \leftarrow \emptyset$; $S \leftarrow \emptyset$; $Q \leftarrow \emptyset$
2: **for** each $u \in V$ **do**
3:    $x.mg \leftarrow computeMG(x)$; $x.it \leftarrow 0$; add $x$ to $Q$
4: **while** $|S| < k$ **do**
5:    $x \leftarrow pop(Q)$
6:    **if** $x.it = |S|$ **then** $S \leftarrow S \cup \{x\}$; $update(x, UC, SC)$
7:    **else**
8:      $x.mg \leftarrow computeMG(x, UC, SC)$;
9:      $x.it \leftarrow |S|$; Reinsert $x$ into $Q$ and heapify

---

Once the first phase is completed, we use the standard greedy algorithm with the CELF optimization [12] to select the seeds (Algorithm 3). The algorithm maintains a queue $Q$ where an entry for a user $x$ is stored in the form $\langle x, mg, it \rangle$, where $mg$ represents the marginal gain of user $x$ with respect to seed set in iteration $it$. $Q$ is always kept sorted in decreasing order of $mg$. Initially, the influence spread of each node is computed and $Q$ is built (lines 2-3). In each iteration, the top element $x$ of $Q$ is analyzed. If $x$ is analyzed before in the current iteration (that is, $x.it = |S|$), then it is picked as the next seed node and the subroutine *update* is called (line 6). On the other hand, if $x.it < |S|$, we recompute the marginal gain of $x$ with respect to $S$ by calling the subroutine *computeMG* (line 8). Then $x.it$ is set appropriately and $x$ is re-inserted into $Q$ (line 9).

---
**Algorithm 4** *computeMG*

**Input:** $x, UC, SC$
**Output:** $mg$
1: $mg = 0$
2: **for** each action $a$ such that $\exists u : UC[x][u][a] > 0$ **do**
3:    $mg_a \leftarrow 1/\mathcal{A}_x$
4:    **for** each user $u$ such that $UC[x][u][a] > 0$ **do**
5:      $mg_a \leftarrow mg_a + UC[x][u][a]/\mathcal{A}_u$
6:    $mg \leftarrow mg + mg_a(1 - SC[x][a])$

---
**Algorithm 5** *update*

**Input:** $x, UC, SC$
1: **for** each action $a$ such that $\exists u : UC[x][u][a] > 0$ **do**
2:    **for** each $u$ such that $UC[x][u][a] > 0$ **do**
3:      **for** each $v$ such that $UC[v][x][a] > 0$ **do**
4:        $UC[v][u][a] \leftarrow UC[v][u][a] - UC[v][x][a] \cdot UC[x][u][a]$
5:      $SC[u][a] \leftarrow SC[u][a] + UC[x][u][a] \cdot (1 - SC[x][a])$

---

Algorithm 4 computes the marginal gain of a node $x$ with respect to the current seed set $S$. It leverages Theorem 3 to do this efficiently. For an action $a$, lines 3-5 compute the term $\sum_{u \in V} \frac{1}{\mathcal{A}_u} \cdot \Gamma_{x,u}^{V-S}(a)$ and line 6 multiplies it by the term $(1 - \Gamma_{S,x}(a))$. Finally, whenever a user $x$ is added to the seed set, the subroutine *update* is invoked and Algorithm 5 updates both $UC$ and $SC$, using Lemmas 2 and 3.

**Memory requirements:** Our algorithm requires to maintain the data structure $UC$ whose size is potentially of the order of $\sum_{a \in \mathcal{A}} |V(a)|^2$. In reality, total credit decreases sharply with the length of paths. Thus by ignoring values that are below a given truncation threshold $\lambda$, the memory usage by our algorithm can be kept reasonable. We study the effect of $\lambda$ in the experiments in the next section.

## 6. EXPERIMENTAL EVALUATION

The goals of our experiments are manifold. At a high level, we want to evaluate the different models and the optimization algorithms based on them with respect to accuracy of spread prediction, quality of seed selection, running time, and scalability. We perform additional experiments on the CD model and on the influence maximization algorithm based on it, to explore the impact of training data size on the quality of the solution and the impact of truncation threshold on the quality, running time, and memory usage. The source of the code used in our experiments is available at http://people.cs.ubc.ca/~goyal/code-release.php.

We experiment on the same two real world datasets of Section 3 (Table 1). While we use the "large" versions of the datasets only to study the scalability of our method, "small" versions of the datasets are used to compare our algorithm with other methods (that do not scale to the large versions).

**Methods Compared.** Since methods based on arbitrarily assigning edge probabilities are dominated by those that learn them from the past propagation traces (see Section 3), in our evaluation, we focus only on the following models.

**IC model** with edge probabilities learnt from the training set by means of the EM method [14]. In all the experiments we run 10k MC simulations.

**LT model** with 10k MC simulations. We take ideas from [10] and [7] and learn weights as $p_{v,u} = A_{v2u}/N$ where $A_{v2u}$ is the number of actions propagated from $v$ to $u$ in the training set and $N$ is the normalization factor to ensure the sum of incoming weights on each node is 1.

**CD model** with direct credit assigned as described in Equation (9). Unless otherwise mentioned, the truncation threshold $\lambda$ is set to 0.001 (see section 5.3). Later we also study effect of different truncation thresholds.

**Accuracy of Spread Prediction.** In Section 3 ("Experiment 2"), we evaluated methods using IC and LT models where edge probabilities are arbitrarily assigned and methods that learn them from available data, with respect to the accuracy of spread prediction. We conduct a similar experiment to compare the IC, LT, and CD models. Fig. 3 shows the RMSE (computed exactly in the same way as in Section 3) in the spread predicted by the IC, LT, and CD models, as a function of actual spread for both datasets. An interesting observation from the figure is that while IC beats LT by a large margin on FLIXSTER_SMALL, it loses to LT on FLICKR_SMALL by a considerable margin. On the other hand, the CD model performs very well on both the datasets.

In order to have a better understanding of the results, we conduct a detailed analysis. Fig. 4 depicts the proportion of propagation traces captured within a given absolute error, which is the absolute difference between the estimated spread and actual spread. More precisely, for a given method, a point $(x, y)$ on its plot says that the fraction of propagation traces (in the test set) on which the (absolute) prediction error of that method is $\leq x$, is $y$. For instance, on FLIXSTER_SMALL, for absolute error $\leq 30$, CD model captures 67% of propagations (that is, 3391 out of 5128 propagations). On the other hand, the percentages of propagations captured within the same error by IC and LT model are 46% and 26% respectively. Once again, it can be seen that while IC performs better than LT on FLIXSTER_SMALL, it's the other way round on FLICKR_SMALL. This plot shows conclusively that within any given error tolerance, CD is able to capture a much higher fraction of propagation traces than IC and LT, on both data sets, confirming that CD model is more accurate when it comes to predicting the influence spread of a given seed set.

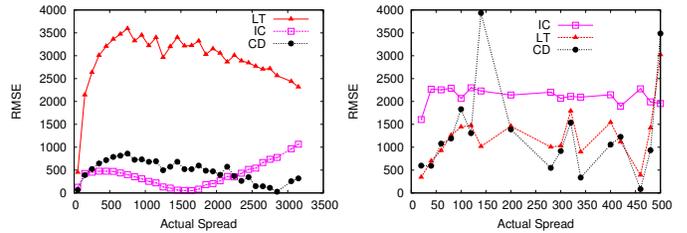

**Figure 3:** RMSE vs Propagation Size on Flixster_Small (left) and Flickr_Small (right).

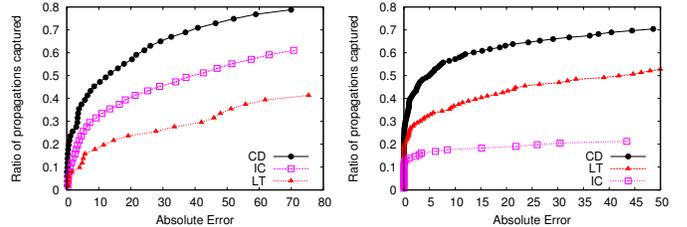

**Figure 4:** Number of propagations captured against Absolute Error on Flixster_Small (left) and Flickr_Small (right).

**Seed Set Intersection.** Having established that CD is much more accurate in predicting actual spread, we next examine the question, how close to each other are the (near) optimal seed sets for the influence maximization problem, obtained by running the greedy algorithm under different models. Fig. 5 shows that the intersection of seed sets obtained from IC model with the seed sets obtained from LT and CD models is empty. On the other hand, there is a significant (~50%) overlap between CD and LT models. We note since the greedy algorithm with MC simulations runs too slow on FLICKR_SMALL (more on this later), in Fig. 5 we use PMIA [2] (for IC model) and LDAG [4] (for LT model) heuristics to obtain the seed set (only for FLICKR_SMALL) in order to finish this experiment within a reasonable time.[5] This shows that the seed sets obtained from IC and LT models are very different from CD model. The difference is much more pronounced in the case of IC model. These findings are significant since together with the results of the previous experiment, they offer some evidence that the seeds chosen by IC and LT models run the risk of being poor with respect to the actual spread they achieve. We strengthen this evidence by conducting the next experiment.

**Spread Achieved.** In this experiment, we compare the influence spread achieved by the seed sets obtained from the three methods. For the sake of completeness, we also include the heuristics High Degree and PageRank which select as seeds the top-$k$ nodes with respect to degree and PageRank score respectively, as done in [10, 2].

---

[5]Chen et al.[2, 4] have shown the spread obtained from PMIA and LDAG are very close to those obtained via MC simulations for IC and LT.

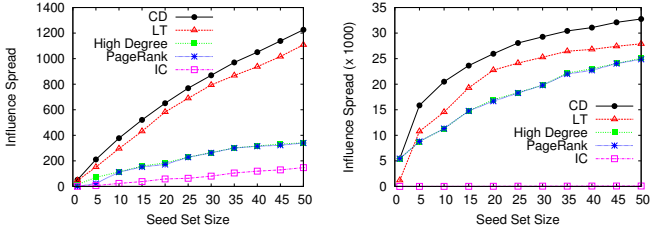

**Figure 5: Size of seed set intersection for $k = 50$ on Flixster_Small (left) and Flickr_Small (right).**

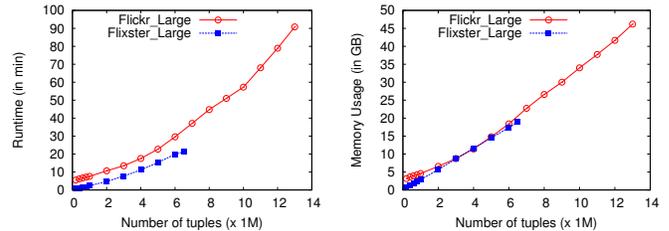

**Figure 6: Influence spread achieved under CD model by seed sets obtained by various models on Flixster_Small (left) and Flickr_Small (right).**

One issue we face is that due to the sparsity issue, we cannot determine the actual spread of an arbitrary seed set from the available data. The next best thing we can do is pick a model that enjoys the least error in spread prediction and treat the spread predicted by it as actual spread. In this way, for any given seed set, we can use that model to tell us (its best estimate of) the actual spread. Given that CD model is found to be closest to reality in predicting the spread of a seed set (see Fig. 3 and 4), we use the spread predicted by it as the actual spread. The results of this experiment, depicted in Fig. 6, confirm that on both data sets the spread achieved by the seed sets found by methods using the IC and LT models falls far short of the actual spread, which is best approximated using the CD model. A surprising observation is that IC model performs poorly, even worse than heuristics like High Degree and PageRank. We looked in the data and found that the seeds picked by IC model are nodes which perform a very small number of actions, often just one action, and should not be considered as high influential nodes. We investigate the reasons below.

For instance, on FLIXSTER_SMALL, the first seed picked by the IC model is the user with Id 168766. While its influence spread under IC model is 499.6, it is only 1.08 under CD model. In the data, the user 168766 performs only one action and this action propagates to 20 of its neighbors. As a result, the EM method [14] ends up assigning probability 1.0 to the edges from 168766 to all its 20 neighbors, making it a high influence node, so much that it is picked as the first seed. Obviously, in reality, 168766 cannot be considered as a highly influential node since its influence is not statistically significant. In an analogy with Association Rules, the influence of user 168766 can be seen as a maximum *confidence* rule, but which occurs only once (absolute *support* = 1).

A deeper analysis tells us that most of the seeds picked by the IC model are of this kind: not very active nodes that, in the few cases they perform an action, do have some followers. We checked and found that the average number of actions performed by these seeds is 30.3 (against the global average of 167). On the other hand, the average number of actions performed by seed set obtained from our CD model is 1108.7. We found a similar behavior in FLICKR_SMALL.

**Running Time.** In this experiment, we first show results on the small versions of the data sets, for all three models, as a function of number of seed nodes selected. All the experiments are run on an Intel(R) Xeon(R) CPU X5570 @ 2.93GHz machine with 64GB RAM running Opensuse 11.3. The algorithms are implemented in C++.

Fig. 7 reports the time taken (in minutes, on log scale) by the various models. It can be seen that our method is several orders of magnitude faster. For instance, to select 50 seeds on FLIXSTER_SMALL, while the greedy algorithm (with CELF optimization) takes 40 and 25 hours under IC and LT model respectively, our algorithm takes only 3 minutes.

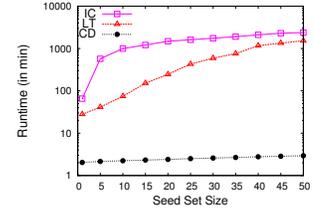

**Figure 7: Running Time Comparison.**

We do not show a similar plot for FLICKR_SMALL as the experiment takes too long to complete (for IC and LT models). At the time of writing this paper, while the experiment for IC model ran for 27 days without even selecting a single seed, the experiment for LT model took the same time to pick only 17 seeds. On the other hand, our algorithm takes only 6 minutes to pick 50 seeds.

**Figure 8: Runtime (left) and memory usage (right) against number of tuples.**

**Scalability.** Next, we show the scalability of our algorithm with respect to the size of the action log, in number of tuples. For this purpose, we created the training data set by randomly choosing propagation traces from the complete action log and selecting all the corresponding action log tuples. In Fig. 8 and 9, the $x$-axis corresponds to the number of tuples in the training set.

Fig. 8 (left) shows the time taken by our algorithm to select 50 seeds against the number of tuples used. It should be noted that most of the time taken by our algorithm is consumed in scanning the action log. For example, it takes 15 minutes to select the seed set when 5M tuples are used on FLIXSTER_LARGE, out of which, 11.6 minutes are spent on scanning the action log and only 3.4 minutes are incurred in selecting the seed set.

Fig. 8 (right) presents the memory usage with respect to the number of action log tuples used to select the seed set of size 50. Our algorithm's memory usage is proportional to the number of training tuples used: on FLIXSTER_LARGE using 6.5M tuples, it requires approximately 16GB, while on 13M tuples on FLICKR_LARGE, it requires approximately 46GB. This raises the question *how much training data is*

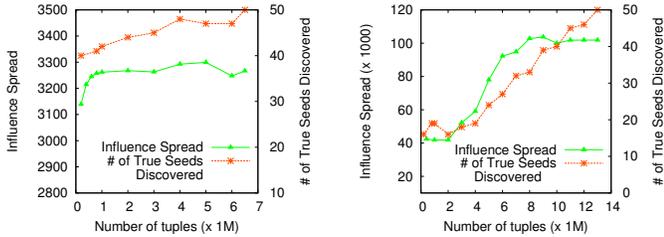

**Figure 9: Influence spread achieved and number of "true" seeds with respect to number of tuples used on Flixster_Large (left) and Flickr_Large (right).**

| $\lambda$ | Influence Spread | True seeds discovered | Memory usage (GB) | Runtime (in min) |
|---|---|---|---|---|
| 0.1 | 2959 | 38 | 2.1 | 5.25 |
| 0.01 | 3220 | 45 | 6 | 8.62 |
| **0.001** | **3267** | **48** | **18.8** | **21.25** |
| 0.0005 | 3267 | 49 | 26 | 25.9 |
| 0.0001 | 3270 | 50 | 51 | 46.7 |

**Table 4: Effect of truncation threshold $\lambda$ on Flixster_Large. "True seeds" are the ones obtained when $\lambda = 0.0001$.**

needed to select a good seed set, which we study next.

**Effect of Training Data Size.** Fig. 9 shows the convergence of the output of our algorithm with respect to number of tuples used to select the seeds. Both plots have a double $y$-axis (left and right). On the left side, we have the spread of influence by the seed set obtained using a sample of training data, while on the right side, we have the overlap of the seed set found with the "true seeds", i.e., the seeds selected by using the complete action logs, i.e., all 6.5M tuples in Flixster_Large and all 13M tuples in Flickr_Large.

As can be seen, the quality of the seed set obtained by using only 1M tuples is as good as using all 6.5M tuples in case of Flixster_Large. Similarly, in Flickr_Large, the influence spread "converges" after 8M tuples.

These observations suggest that we need to use only a small sample of the propagation traces (or action log) to select the seed set and as a result, even though our algorithm can in principle be memory intensive when the action log is huge, in reality, the memory requirements are not that high.

**Effect of truncation threshold.** Finally, we show the effect of truncation threshold $\lambda$ on the accuracy, memory usage and running time of our algorithm in Table 4. As expected, as we decrease the truncation threshold, while accuracy (measured in terms of number of "true" seeds discovered and influence spread achieved) improves, memory requirement and running time increase. Both the influence spread and "true seeds discovered" essentially saturate at $\lambda = 0.001$. Note that in all our previous experiments, we used $\lambda = 0.001$ which is a good choice as can be seen from the table. The results on Flickr_Large and on small versions of the datasets are similar.

## 7. CONCLUSIONS AND DISCUSSION

While most of the literature on influence maximization has focused mainly on the social graph structure, in this paper we proposed a novel data-based approach, that directly leverages available traces of past propagations.

Our Credit Distribution model directly estimates influence spread by exploiting historical data, thus avoiding the need for learning influence probabilities, and more importantly, avoiding costly Monte Carlo simulations, the standard way to estimate influence spread. Based on this, we developed an efficient algorithm for influence maximization. We demonstrated the accuracy on real data sets by showing the CD model by far is closest to ground truth. We also showed that our algorithm is highly scalable.

Beyond the main contributions, this paper achieves several side-contributions: (1) Methods which arbitrarily assign influence probabilities suffer from large error in their spread prediction compared with those that learn these probabilities from data. (2) The former methods end up choosing seed sets very different from the latter ones, suggesting the seeds they recommend may well have a poor spread. (3) The greedy algorithm using learned influence probabilities is robust against some noise in the probability learning step. (4) The IC and LT models, using learned influence probabilities, choose seed sets very different from each other, and in turn different from the CD model, which is by far closest to ground truth. These observations further highlight the need for devising techniques and benchmarks for comparing different influence models and the associated influence maximization methods.

**Acknowledgments.** This research was partially supported by a strategic grant from NSERC Canada on Business Intelligence Network, and a grant from the Spanish Center for the Development of Industrial Technology under the CENIT program, project CEN- 20101037, Social Media. Thanks to Jamali and Ester for sharing the Flixster dataset [9].